\documentclass[conference]{IEEEtran}
\IEEEoverridecommandlockouts
\usepackage{cite}
\usepackage{amsmath,amssymb,amsfonts}
\usepackage{algorithmic}
\usepackage[hyphens]{url}
\usepackage{graphicx}
\usepackage{braket}
\usepackage{quantikz}
\usepackage{hyperref}
\usepackage{textcomp}
\usepackage{xcolor}
\def\BibTeX{{\rm B\kern-.05em{\sc i\kern-.025em b}\kern-.08em
    T\kern-.1667em\lower.7ex\hbox{E}\kern-.125emX}}
\begin{document}

\title{The Impact of Feature Embedding Placement in the Ansatz of a Quantum Kernel in QSVMs\\
\thanks{This work was funded by the Business Finland Quantum Computing Campaign, project FrameQ.}
}

\author{\IEEEauthorblockN{Ilmo Salmenperä}
\IEEEauthorblockA{\textit{Computer Science} \\
\textit{University of Helsinki}\\
Helsinki, Finland \\
ilmo.salmenpera@helsinki.fi}
\and
\IEEEauthorblockN{Ilmars Kuhtarskis}
\IEEEauthorblockA{\textit{Computer Science} \\
\textit{University of Helsinki}\\
Helsinki, Finland \\
ilmars.kuhtarskis@helsinki.fi}
\and
\IEEEauthorblockN{Arianne Meijer - van de Griend}
\IEEEauthorblockA{\textit{Computer Science} \\
\textit{University of Helsinki}\\
Helsinki, Finland \\
arianne.vandegriend@helsinki.fi}
\and
\IEEEauthorblockN{Jukka K. Nurminen}
\IEEEauthorblockA{\textit{Computer Science} \\
\textit{University of Helsinki}\\
Helsinki, Finland \\
jukka.k.nurminen@helsinki.fi}
}

\maketitle

\begin{abstract}
Designing a useful feature map for a quantum kernel is a critical task when attempting to achieve an advantage over classical machine learning models. The choice of circuit architecture, i.e. how feature-dependent gates should be interwoven with other gates is a relatively unexplored problem and becomes very important when using a model of quantum kernels called Quantum Embedding Kernels (QEK). We study and categorize various architectural patterns in QEKs and show that existing architectural styles do not behave as the literature supposes. We also produce a novel alternative architecture based on the old ones and show that it performs equally well while containing fewer gates than its older counterparts.
\end{abstract}


\section{Introduction}
Kernel methods enable Support Vector Machines (SVMs) to classify data by projecting it into a higher-dimensional space using a feature map making the data linearly separable~\cite{theodoridis2006pattern}. This linkage allows SVMs to handle complex and nonlinear relationships within the data. While classical kernel methods have been explored extensively, the field of quantum machine learning (QML) has produced a new way of creating them in the form of quantum kernels~\cite{mengoni2019kernel, schuld19, goto21}.

Quantum computers enhance kernel methods by utilizing the exponential size of the Hilbert space to separate data points coming from different probability distributions~\cite{goto21}. This capability allows for the extraction of features from data and the identification of complex relationships with a depth and efficiency not achievable with classical computing~\cite{Kubler21}. While there exists theoretical and experimental evidence for the usefulness of these models, there are still some concerns over their applicability and whether they actually provide an advantage over classical models~\cite{bowles2024better}.

An interesting variant of quantum kernels is Quantum Embedding Kernel (QEK), which takes inspiration from variational circuits, which in addition to the quantum feature map, employ parameterized layers in their architecture~\cite{Hubregtsen22}. These parameterized layers are trained using a process called kernel target alignment, which aims to create a feature map that projects data points with different labels into separate regions in the feature space~\cite{cristianini2001kernel}. The advantage of this approach is that the resulting feature map is more likely to project the data points into linearly separable spaces.

While this method of using quantum kernels has shown some advantages over their more conventional counterparts, such as the capability to tweak the kernel itself to perform better in a given task, there are still some unaddressed issues relating to the structure of the kernel's parameterized parts and the feature maps. In existing literature, there are two architectural styles for parameterized kernel methods that we are interested in: (1) data-first architectures where the feature dependant parts of the kernel come before the parameterized layers~\cite{Hubregtsen22, rodriguezgrasa2024training, mostafa23} and (2) data-last architectures, where the parameterized layers come before feature-dependant layers~\cite{paine23, miroszewski2023detecting}.

While the data-first architectures were proposed by the original research papers about QEKs, some newer articles have separately noted that this style of layering will result in the last of the parameterized layers vanishing due to being right next to its own adjoint in the circuit~\cite{paine23}. This phenomenon will be referred to as the gate erasure bug in this article. The proposed fix to this issue is the data-last architecture, but the performance of these two architectures has not been compared with one another.

In this study, we show that the order in which parameterized layers are interweaved with feature-dependent layers has a considerable impact on the performance and in the worst case can cause the gate erasure bug to occur. While this gate erasure bug is not a widespread issue in the field, we found multiple instances of these worst case architectures in the quantum kernel related publications and very little discussion on the effect itself. In addition, we found out that when scaling up the layer count of the model, the data-last architecture does not outperform the data-first architecture, even if the data-last architecture effectively has one additional parameterized layer.

Due to this finding, we propose a third way of creating QEKs: data-weaved kernels, where the parameterized layers are weaved between feature-dependent layers and show that these styles of kernels outperform their older counterparts in most benchmarking tests. We should note that this study does not indicate anything new about the usefulness of QEKs in general, as the advantages of the new architecture materialize mostly in lower gate counts or possible speed-ups in the optimization process. These are still important metrics, when trying to improve our current-generation QML models to achieve any type of an advantage over the classical ML models. 

The key contributions of this article are as follows:
\begin{itemize}
\item Exploration of the background and theory behind the two existing architectural styles of quantum kernels and introduction of data-weaved kernels (Sections 2 and 3)  
\item Implementation and evaluation of the three different architectural styles on four different datasets (Section 4)
\item Analysis of the influence of architectural style to quantum embedding kernels across various metrics and further discussion of the practical implications they have for quantum kernels in general (Section 5)
\end{itemize} 

\section{Background}

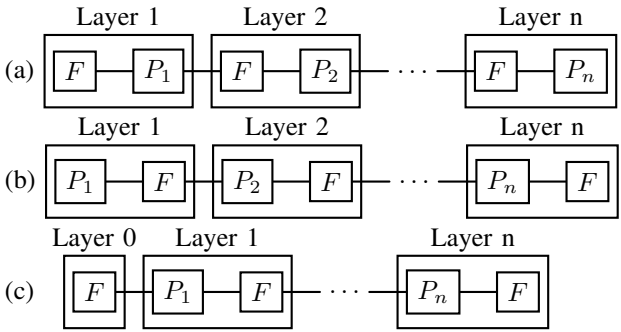
\begin{figure}
    (a)
    \begin{quantikz}
        \gate{F}\gategroup[1,steps=2,style={inner xsep=0pt, inner ysep=2pt}]{Layer 1} & \gate{P_1} & \gate{F}\gategroup[1,steps=2,style={inner xsep=0pt, inner ysep=2pt}]{Layer 2} &\gate{P_2} & \ \ldots\ & \gate{F}\gategroup[1,steps=2,style={inner xsep=0pt, inner ysep=2pt}]{Layer n} & \gate{P_n}
    \end{quantikz}
    \newline
    (b)
    \begin{quantikz}
        \gate{P_1}\gategroup[1,steps=2,style={inner xsep=0pt, inner ysep=2pt}]{Layer 1} & \gate{F} & \gate{P_2}\gategroup[1,steps=2,style={inner xsep=0pt, inner ysep=2pt}]{Layer 2} &\gate{F} & \ \ldots\ & \gate{P_n}\gategroup[1,steps=2,style={inner xsep=0pt, inner ysep=2pt}]{Layer n} & \gate{F}
    \end{quantikz}
    \newline
    (c)
    \begin{quantikz}
        \gate{F}\gategroup[1,steps=1,style={inner xsep=0pt, inner ysep=2pt}]{Layer 0} & \gate{P_1}\gategroup[1,steps=2,style={inner xsep=0pt, inner ysep=2pt}]{Layer 1} &\gate{F} & \ \ldots\ & \gate{P_{n}}\gategroup[1,steps=2,style={inner xsep=0pt, inner ysep=2pt}]{Layer n} & \gate{F}
    \end{quantikz}
    \newline
    \caption{Three different architectural solutions: (a) data-first structure, (b) data-last structure, (c) the data-weaved structure. $F$ is feature-dependant layer and $P_n$ is the n:th parameterized layer. Note that the data-weaved structure has one extra feature-dependent layer before the other layers. }
    \label{fig:architecture_types}
\end{figure}

In this study we concentrate on a specific implementation of quantum kernels called QEKs, and how different layering strategies when designing the overall structure of their feature map affects their performance. One especially interesting aspect of the architectures is how the feature-dependent layers and parameterized layers are interwoven. There are two styles of architectures present in the literature: (1) data-first architecture, where the feature-dependent layers occur before the parameterized layers~\cite{Hubregtsen22, rodriguezgrasa2024training, mostafa23} and (2) data-last architecture, where this order is reversed ~\cite{paine23, miroszewski2023detecting}. Examples of the data-first and data-last architecture can be seen in~\autoref{fig:architecture_types}.

The use of data-last architecture in QEKs is motivated probably due to the gate erasure bug in the data-first architectures, but to our knowledge, there have been no studies done on how having a parameter-layer before any feature-dependent operations affect the performance of a QEK model when compared against the data-first model. Our aim in this study is to investigate how the performance of these models compare against each other to see whether the use of data-last architecture actually makes sense beyond the theoretical aspects and also see whether better models could be created based on these results.

It is also important to note that QEKs are not the only kind of quantum kernel methods that can be impacted by these kinds of choices in architecture. For example~\cite{torobian23}  proposes a way to create kernels through an adaptive process of adding various entangling gates and parameterized gates layer by layer after the feature-dependent layer, where only one of the five types of different gates contains feature-dependant variables. This means that if the model generates a circuit with some non-parameterized parts at the end, the gate erasure bug will cause them to vanish when evaluating values for the kernel matrix.

We would also like to take notice of an article featuring a quantum kernel model that does not strictly refer to itself as QEK, but follows a structure quite similar to theirs~\cite{Glick2024}. In this article a parameterized layer is used to create a preliminary state to which the feature-dependent layer is applied. While this has been shown to increase the performance of the SVM in certain contexts, it is also noted that there exists certain preliminary states which can also harm the performance of the model instead. This result will be relevant during the analysis of the data-last architecture. 

\section{Methodology}

\subsection{Quantum Embedding Kernels}
Kernel methods are a powerful class of algorithms in machine learning, particularly suited for tasks where linear classification methods, such as Support Vector Machines (SVM), face challenges due to non-linearly separable data~\cite{CERVANTES2020}. These methods operate by projecting the original data points $x$ to a higher-dimensional feature space with a feature map $\Phi(x)$, where linear separation becomes feasible. 

Interestingly it is actually not necessary to compute where these projected data points are in the higher-dimensional space, which can be computationally very expensive, or even impossible~\cite{campbell2001}. This is made possible through the kernel trick, which can be used to train the SVM using a similarity measure based on the inner products between data points in the training set~\cite{Scholkopf2000TheKT}. If our feature map has a well defined inner product, we can create a kernel function:

\begin{align}
\label{math1}
    \kappa(x, x') = \braket{\Phi(x),\Phi(x')}
\end{align}

This kernel function can be used to create a kernel matrix K for a set of data points in our training set $D$, where each element of the matrix is the value of the kernel function between two data points. This allows us to find the weights for the SVM inside the projected space, which can then be used to classify or group new data points.

Interestingly the concept of kernel function translates very naturally to quantum computing. If we create a quantum kernel function $\ket{\Phi(x)} = U(x) \ket{0}$ for some data point, the inner product between two different projected data points can be computed easily using the Loschmidt echo test~\cite{kusumoto2021} or the swap test~\cite{blank2020}.

In the Loschmidt echo test, the inner product between two projected data points $x$ and $x'$ is computed by first applying the feature map $U(x)$, followed by the application of the adjoint of the same feature map for a different data point $U(x')^{\dagger}$. The probability of the system being in the initial state, denoted as $|0\rangle^{\otimes n}$, where $n$ is the number of qubits in the system, can be computed as:

\begin{align}
    P(\ket{0}^{\otimes n}) &= \braket{0^{\otimes n} | U(x)^\dagger | 0^{\otimes n}} \braket{0^{\otimes n} | U(x') | 0^{\otimes n}} \\
    &= \braket{0^{\otimes n} | U(x)^\dagger U(x') | 0^{\otimes n}} \\
    &= \braket{\Phi(x) | \Phi(x')}
\end{align}

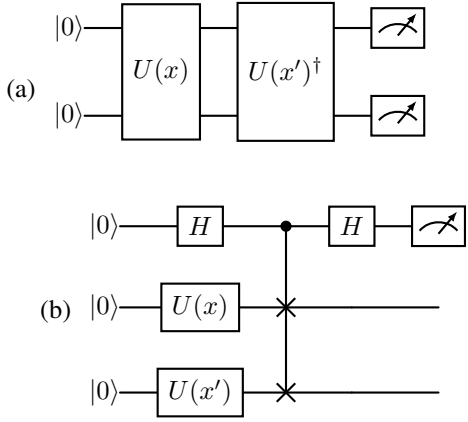
\begin{figure}
    \centering
    (a)
    \begin{quantikz}
        \ket{0} & \gate[2]{U(x)} &\gate[2]{U(x')^\dagger} & \meter{}\\
        \ket{0} & & & \meter{}\\
    \end{quantikz}
    \newline
    (b)
    \begin{quantikz}
        \ket{0} & \gate{H} & \ctrl{2} & \gate{H} & \meter{}\\
        \ket{0} & \gate{U(x)} & \targX{0} & & \\
        \ket{0} & \gate{U(x')} & \targX{0} & & 
    \end{quantikz}
    \caption{Two methods for computing the inner product of the kernel: (a) Loschmidt echo test and (b) Swap test}
    \label{fig:innerp-methods}
\end{figure}

The swap test computes the same inner product, but uses an ancillary qubit and two registers each prepared in one of the states provided by the feature maps $U(x)$ and $U(x')$~\cite{blank2020}. Using two Hadamard operations and a controlled-swap gate in the pattern shown in \autoref{fig:innerp-methods} the probability of measuring the ancilla qubit is given by: 

\begin{align}
P(\text{Ancilla is } 0) &= \frac{1}{2} + \frac{1}{2} |\braket{0^{\otimes n} |U(x)^\dagger U(x') | 0^{\otimes n}}|^2 \\
|\braket{\Phi(x)|\Phi(x')}| &= \sqrt{2 P(\text{Ancilla is } 0) - 1}
\end{align}

Both of these methods end up evaluating the same inner product between two feature-maps, the only difference being the amount of qubits and the circuit depth required by the method. The Loschmidt echo test seems to be currently more popular in the field of research, as qubit counts are a larger bottleneck, but the swap test is equally viable as an approach to implement a quantum kernel. 

Similarly to classical kernels~\cite{Chaudhuri2008}, one issue with quantum kernels is that it is very hard to know what kind of a feature map should be used to solve a certain problem~\cite{schuld19}. Especially when solving difficult problems, we cannot guess beforehand the correct feature map to use to make the data set linearly separable, which means that a lot of time needs to be spent on trying different kinds of kernel functions to find one that allows us to solve the problem using a SVM. To combat this phenomenon, Quantum Embedding Kernel (QEK) model was developed, which combines quantum kernels with the style of learning from variational quantum circuits.

QEKs are a special case of quantum kernels, where in addition to the feature encoding layer of $U_1(x)$, the circuit also has a parameterized layer $U_2(\theta)$~\cite{Hubregtsen22}. The conventional wisdom from quantum neural networks is to interweave these parameterized layers with feature encoding layers to increase the expressiveness of the model~\cite{du2020}. An interesting feature of QEKs is that the parameterized parts of the feature map allow us to optimize our feature map to be more likely to project our distinctly labeled data points into separate spaces, using a process called kernel-target alignment.

Kernel-target alignment aims to fine-tune the quantum kernel's parameters to achieve an optimal representation of the data~\cite{cristianini2001kernel}. In this method, we construct an ``ideal" kernel matrix $K_{ideal}$ from a set of training data points, where each value is based on whether the labels are the same or not:

\begin{align}
K_{ideal}(x, x') = y(x) y(x')
\end{align}

where $y(x)$ gives us whether the label for data point $x$ is a $-1$ or $1$.

Using this ideal kernel, we can construct a metric called target alignment, which we can use to evaluate similarity between the ideal kernel and the kernel we have at hand~\cite{cristianini2001kernel}:

\begin{align}
A(K, K_{\text{ideal}}) &= \frac{\braket{K, K_\text{ideal}}_F}{\sqrt{\braket{K, K}_F\braket{K_\text{ideal}, K_\text{ideal}}}_F} \\
&= \frac{\sum_{i,j} y(x_i) y(x_j) k(x_i, x_j) }{n \sqrt{\sum_{i,j} k(x_i, x_j)^2}}
\end{align}

Where $\braket{\cdot, \cdot}_F$ is the Frobenius inner product, $n$ is the amount of samples from the training set $D = {x_1, ..., x_n}$ that we use to compute the target alignment metric. 

The core concept of target alignment is that if we can tweak the behaviour of $K$ using parameterized gates, we can optimize the performance of kernel function itself to become more aligned with this ideally separating kernel. Studies have shown that target alignment is able to improve the performance of kernel methods in learning tasks~\cite{wang2015overview} and which also means that instead of having to choose the correct kernel function for a given task, we can learn the feature map based on the problem itself. 

\subsection{Quantum kernel architectures}

Interesting question related to QEKs is how we should interweave the parameterized layers with the feature dependent layers. Like we showed previously in the background section of this article, there are two main approaches of constructing the kernel layering structure: (1) data-first architectures~\cite{Hubregtsen22} and (2) data-last architectures~\cite{paine23}. 

In the data-first architectures the feature dependent layers are put before the parameterized layers, as shown in \autoref{fig:architecture_types}. This way of constructing QEKs  suffers from the gate erasure bug, which causes the architecture to have redundant gates at the end of the circuit.

The gate erasure bug occurs, when the inner product produced by the quantum circuits has data-independent gate operations at the end of the model circuit, meaning we can separate the unitary operator into two parts $U(x, \theta)\ket{0} = U_2(\theta) U_1(x)\ket{0}$, where $U_1(x)$ is the part of the feature map which is dependant on features and $U_2(\theta)$ is dependant only on parameters of the feature map $\theta$. This causes the inner product to erase these operations completely from the inner product and, by extension, from the circuit:

\begin{align}
\kappa(x, x') &= \braket{0|U(x', \theta)U(x, \theta)|0}\\
&= \braket{0| U_1(x')^\dagger U_2(\theta)^\dagger U_2(\theta) U_1(x)|0} \\
&= \braket{0| U_1(x')^\dagger U_1(x)|0} 
\end{align}

This effect is immediately apparent when utilizing the Loschmidt echo test to evaluate the inner product - the constant gate operations inside parameterized parts of the circuit are positioned adjacent to their adjoint counterparts, resulting in their combination into identity operators. This results in all of the parameterized parts after the last feature-dependent gates to be completely irrelevant from the point of view of the distance metrics used to train the SVM. Additional problems will be caused by gates that commute with the feature-dependent gates at the end, which are also effectively cancelled out, as they can always be transferred to either side of the commuting gates.

While this phenomenon is less apparent when used with the swap test method, mathematically the same issue still applies. All gates at the end of the feature map $\ket{\Phi(x)}$  that are not dependent on the feature vector, will always resolve into identity operators, due to the inner product $\braket{\Phi(x)|\Phi(x')}$. This bug is also relevant to quantum kernels outside of QEKs, if they happen to have extensive amounts of data-independent gates after applying the feature-dependent operators, such as CNOTs or Hadamard operators. 

\begin{figure*}[t!]
\centering
\begin{tabular}{c c}
     \includegraphics[scale=0.5]{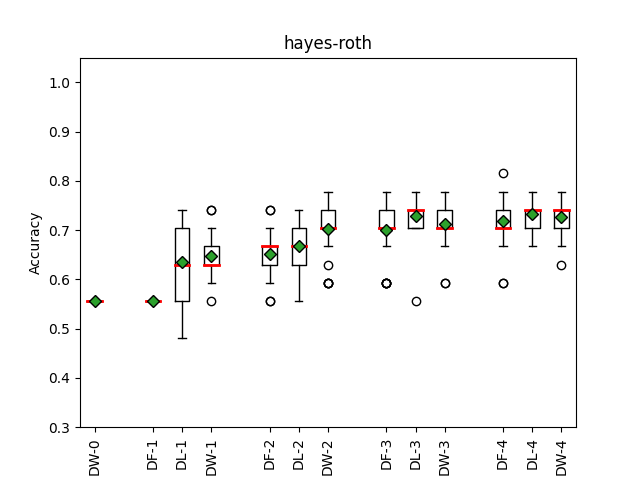} & 
     \includegraphics[scale=0.5]{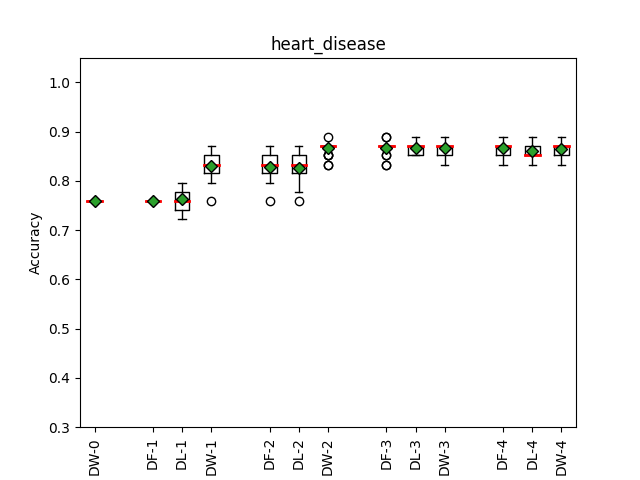} \\
     \includegraphics[scale=0.5]{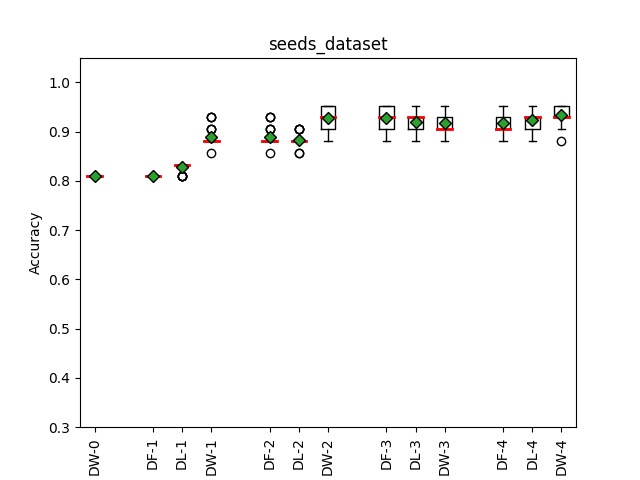} &
     \includegraphics[scale=0.5]{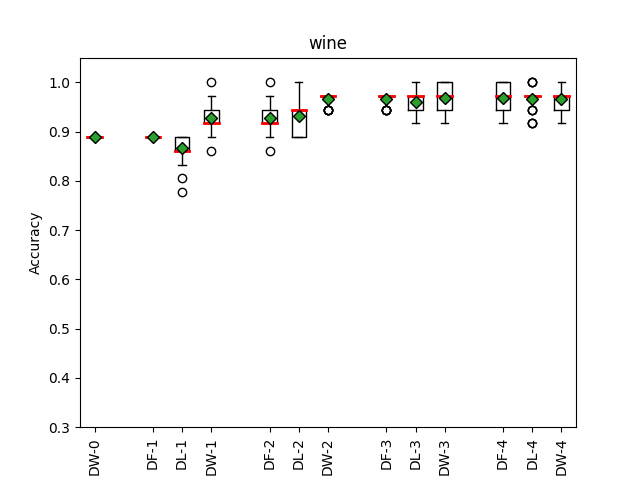} \\
\end{tabular}

\caption{Box plot of test set accuracies of the three kernel architectures on four different datasets. Models are labelled as: DW - Data-weaved, DF - Data-first and DL - Data-last. The green diamond is the mean accuracy of the 25 models, while the red bar is the median accuracy. White dots represent outlier data points in the accuracy. The results are grouped together by the amount of parameter layers in the architecture, indicated by the number after the label. All the results were gained from the same test training split.}
\label{fig:results}
\end{figure*}

The second architecture, the data-last architecture,  was created to address the issue of gate erasure by reversing the order of the parameterized layers and feature encoding layers~\cite{paine23}. While this approach does not suffer from the gates negating each other at the end of the feature map, it is unclear whether having any parameterized parts before feature dependent layers increases the performance of QEKs.

One key reason why there is no guarantee that the data-last kernels would outperform data-first kernels is, that applying a constant rotation before any feature-dependent operations could mostly just affect how the gates in the first data-embedding layer change the state. For example, causing the initial state to be a set of vectors in direction of the X-axis, if the feature-map uses rotations around the X-axis to embed the data into the system, the obvious outcome will be that all the inner products between projected data points will be always 1, as the feature embedding does not change the state at all. By this logic, it is also possible that there exists a preliminary state that could improve the ability of the kernel to separate data points from each other, as a poor choice of embedding rotations directions could ruin the kernel's capability to separate projected data points from each other, but this can of course be facilitated by informed choices when designing a circuit. Preliminary entanglement could provide some useful features to learning process, but there is also very little theoretical reason to believe so. 

We propose a third way of creating quantum embedding kernels: the data-weaved kernels. In this approach the parameterized layers are put in between of data layers, so that there are no unnecessary gates after the feature map, and there are also no preliminary transformations of the quantum state before any feature encoding. Interestingly all data-first kernels are mathematically equivalent to data-weaved kernels with one less parameterized layer at the end, as the gate erasure bug will reduce them to this data-weaved structure. 

In this study we group the three models together according to how many parameter layers they have, as we are interested specifically how different parameterization strategies affect the performance of the models. This means that the data-weaved strategy contains one extra feature-dependent layer compared to equivalent alternative structures. We consider this to be the most sensible way to group the models together, as the feature-dependent layers tend to require less gates compared to parameterized layers, and they do not affect the scaling of the optimization strategy used in the training of the model.

\subsection{Experimental setup}
\label{sec:experimental-setup}
To evaluate the performance and trainability of these three models, we focus on classification accuracy, kernel alignment and time taken across a variety of datasets, including Hayes-Roth~\cite{misc_hayes-roth_44} (3 classes), Heart Disease~\cite{misc_statlog_(heart)_145} (2 classes), Seeds~\cite{misc_seeds_236} (3 classes), and Wine~\cite{misc_wine_109} (3 classes), all features normalized between 0 and 1. The trained kernels were then given to a support vector classifier model (SVC), which was trained to be either a binary classifier in the case of having two classes, or a multi-class classification using a one-vs-many scheme~\cite{hsu2002}. 

In our experiment design, certain parameters remain constant across all trials for consistency: we use 5 qubits (wires), a batch size of 5, run 5000 optimization iterations, and conduct alignment tests every 250 iterations. 25 different models were trained for each variable parameters included the number of layers (ranging from 1 to 5) to assess the impact of these factors on performance. The test-training split was fixed to be consistent for each dataset, and the split was chosen manually from among 25 candidates to ensure that the models are easy to compare to each other. The results are consistent with the random test-training splits, but way less noisy, as the average dataset size was very small and had a large impact of the end performance. For the optimization method, we use gradient descent with the finite differentiation method. The ansatz structure is explained in Appendix A.

\begin{figure*}[t!]
    \centering
    \includegraphics[scale=0.64, trim={10cm 0 10cm 0}]{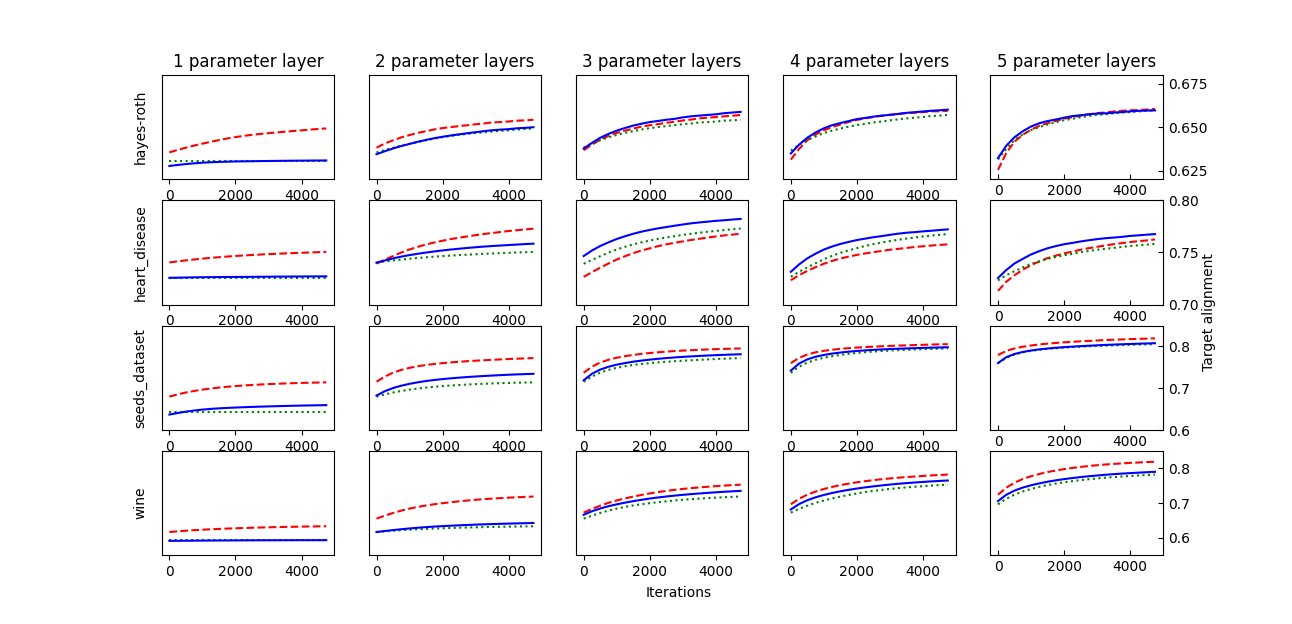}
    \caption{Mean target-alignment values as the function of time for different datasets and layers. Red dashed line represents the data-weaved architecture, green dotted line represents the data-first architecture and blue solid line represents the data-last architecture. The lines are grouped together by the amount of parameter layers in the architecture.}
    \label{fig:alignment}
\end{figure*}

\section{Results}

The impact of the three architectural strategies on accuracy of the trained model can be seen in the Figure \ref{fig:results}. From the results we can see that target alignment does produce models with higher levels of accuracy and the more layers you add to the models, the better the performance tends to be. With most of our datasets the performance seems to peak at 3 parameterized layers, after which adding more layers does not affect the accuracy of the model in any meaningful sense. 

The impact of the gate erasure bug is very visible in these results. The data-first model with one parameterized layer has no variance in results as the single parameterized layer does nothing and almost never outperforms the other two architectures. With two parameterized layers or more the data-first models seems to be performing equally well to the data-last architecture, which is surprising, as the last parameterized layer should still resolve into an identity operator as predicted by the theory, resulting in one less parameterized layer compared to the data-last kernel. This seems to indicate that parameterized layers in between feature-layers tend to be more effective at increasing the accuracy of the trained model.

The performance of the data-last architecture is somewhat consistent with existing results - with one parameterized layer, the data-last architecture can achieve a higher accuracy than the data-first architecture, but sometimes they can also end up performing worse performance in the given task. This speaks to the volatility of the results and shows that there is no guarantee that the introduction of preliminary quantum states would help with separating the data points from one another. 

Data-weaved model achieves higher accuracies in almost all test cases, with an equal amount parameterized gates compared to the other two models. The only exception for this is the Hayes-Roth dataset with one layer of parameters, and even there the average performance becomes the consistent with the other results further down the line. 

Important thing to note about the performance of the three models is that the choice of architecture does not really affect the peak performance of the end model, as long as we increase the layer size enough. This means that the only meaningful difference between the three models is how many gates the models have, and more importantly, how many parameters are needed to optimize when training the model.  

The next interesting thing is how the target-alignment changes during training with the three models, as shown in \autoref{fig:alignment}. One interesting overall result seems to be that more layers in the model seem to increase not only the end alignment of the model, but also the initial alignment. An additional general observation is that higher alignment during training indicates better a performance when cross-referencing results with \autoref{fig:results}.

\begin{figure*}[ht!]
    \centering
    \includegraphics[scale=0.62, trim={10cm 1cm 10cm 0cm}]{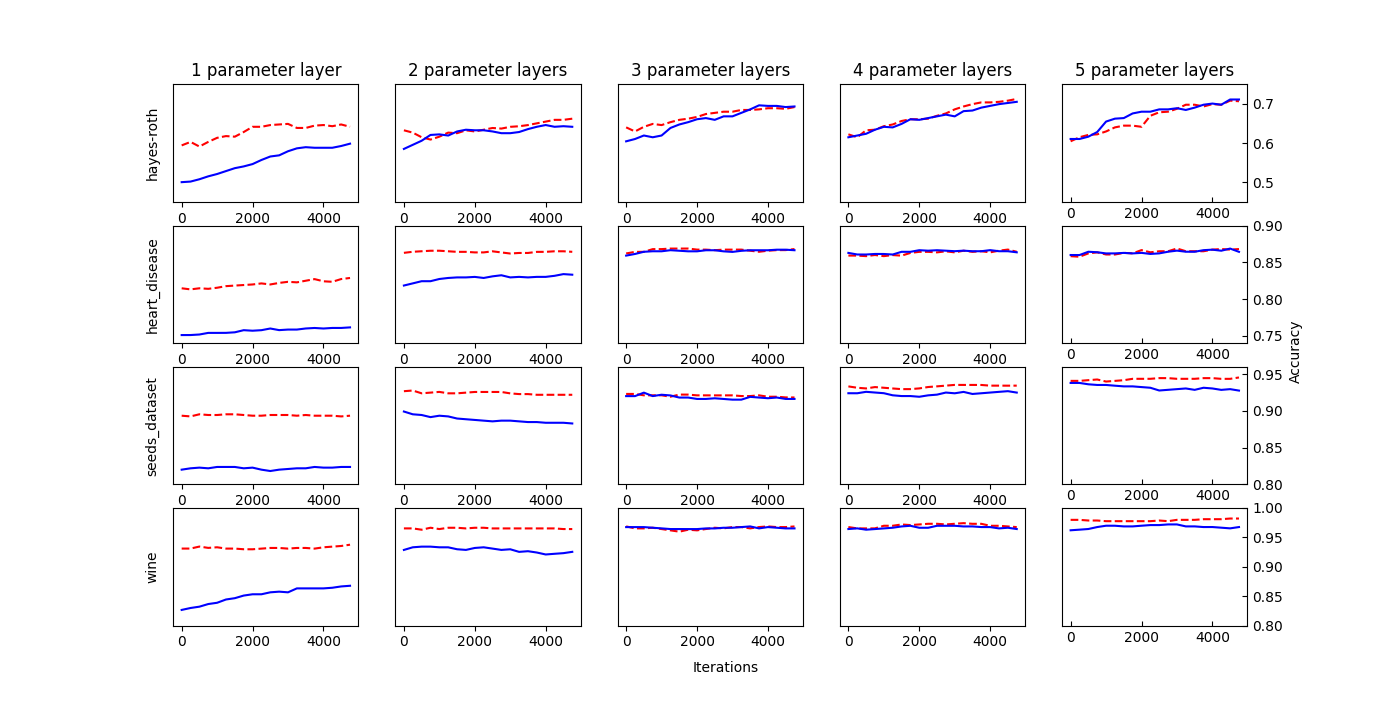}
    \caption{Average test accuracy for 25 different models over training iterations for the data-weaved model (The red dashed line) and the data-last model (blue solid line). The data-first model was omitted due to it being identical to the data-weaved model with one less parameter layer.}
    \label{fig:training}
\end{figure*}

One key observation with these three models from the point of view of alignment is that the data-weaved architecture seems to improve the most during the training process. This seems to indicate that this style of kernels is more trainable than its alternatives. There was one exception to this, which is the heart-disease dataset where the data-last model seems to out-align the data-weaved model, but it has to be also noted that the improvement in alignment does not result in increases in resulting accuracy of the model, as seen in \autoref{fig:training}.

With one parameterized layer the data-first shows no improvement in alignment during training, which is as expected. Another interesting feature is that if we were to overlay the mean alignment of the data-first kernels with the data-weaved kernels with one less parameter layer, we could see that the two graphs are completely identical, reinforcing the idea, that the data-first kernel is in fact a data-weaved kernel with one unnecessary layer at the end of the circuit. 

The third interesting graph is how does the accuracy of the models evolve during training, which can be seen from \autoref{fig:training}. This shows that, on average, the data-weaved models achieve better accuracy with less amounts of parameterized layers on average and some times even better accuracy overall. An interesting thing to note is that in some cases with large amount of layers, the kernel target alignment process does not really improve accuracy of the model that much, or even in some cases, it outright decreases the accuracy of the outputted model. 

\begin{figure}[hb!]
    \centering
    \includegraphics[scale=0.52]{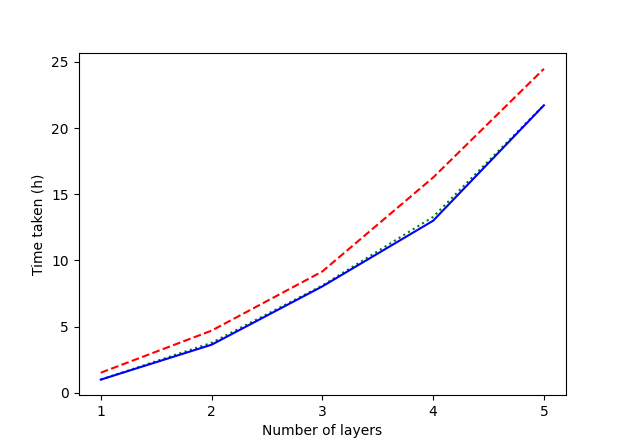}
    \caption{The average simulation time taken in hours for the amount of parameterized layers in the model. The blue solid line is the data-last model, the green dotted line is the data-first model and the red dashed line is the data-weaved model. The data-first model is almost equivalent to the data-last model, which is why it is barely visible in the graph.}
    \label{fig:time}
\end{figure}

The last metric in which these three models are different is the time that it takes to simulate these three models, as shown in \autoref{fig:time}. The results show that while the data-weaved model takes a bit more time to train compared to the data-first and data-last models which contain equal amounts of parameterized layers, as we are able to get better results with less layers, the data-weaved kernels will end up outperforming the alternatives in optimal configurations. This result shows that the use of data-weaved structure should save us classical computational resources in research. While this does not indicate that meaningful amounts of time could be saved when computing these things with real quantum computers, in those cases aspects like circuit length becomes a larger issue.

Interestingly we can also see that the data-first and data-last kernels take equally as long to train, which is an indication that Pennylane does not by default transpile its parameterized circuits into more optimal forms. If this were to be the case, it would mean that the data-first kernels should take as much time as a data-weaved kernel with one less layer, as they are equivalent in resource costs to simulate.

\section{Discussion}

The differences between the three architectural choices seem to be consistent with the concepts from theoretical background. It seems that there are very few reasons to have parameterized circuits before or after any encoding layers, except if we are interested in using only one layer of parameters. While this does not make sense from the point of view of QEKs, there might be other styles of quantum kernel methods that might benefit from this strategy. 

The simulated results show that while all of these models are capable of producing acceptable results, the data-weaved kernels are usually able to do more with similar access to resources. This means that according to these results, using data-weaved kernels seems to be the most sensible choice out of the three possibilities. Further research could be made into how much different choices of ansatze could impact the results, but at least in our experiences using different choice of feature-encoding or parameterized layers did not effect the results in meaningful ways.

The key reason for why data-weaved architectures should be used with quantum kernels comes mostly down to resource usage, at least in simulated environments. They are not more accurate than their alternative counterparts, as the user is free to choose how many layers they want to use in the training process of the model. The advantage gained from using data-weaved kernels is very dependent on the problem in question, but for example using our ansatze and feature-embedding layers, for a dataset with ten features, the decrease in the amount of gates would be around $~11\%$ with one qubit gates (from 90 to 80) and $~30\%$ with two qubit gates (from 30 to 20), which can be a very meaningful difference in the world of high performance computing. These savings would increase even further if more complicated ansatze are used.

An additional key factor influencing these results from classical point of view is the performance of optimizers in handling non-contributing parameters. In cases where parameters are nullified by the gate erasure bug, an optimizer not designed to disregard these parameters may experience inefficiencies. This slowdown arises as the optimizer attempts to adjust parameters that have no impact on the outcome. For example, optimizers using the parameter shift rule would probably spend some amount of time evaluating parameters that do not have an effect to the output system~\cite{Wierichs2022generalparameter}, while the use of SPSA-optimizers would cause the unnecessary parameters to be negligible during training, as they adjust all the parameters at the same time~\cite{spall1998implementation}. 

From the quantum computing point of view, the choice of architecture can have a large impact on the performance of the circuit. If the data-weaved kernel requires one less parameterized layer of gates to operate at the optimal level of accuracy, this could have a sizable difference in performance compared to the data-last kernels, as at least in the near term future we have to be mindful of device noise caused by additional gate operations. Additional noise can be very harmful from the point of view of training a QML model, as shot noise is heavily linked with trainability of the model in question~\cite{Cerezo2022}. This problem is exacerbated further by the use of data-first kernels, which contain a layer of unnecessary gates in the end, that will be in the best case optimized out from the circuit, and in worst case, kept by the transpiler that refuses to remove the unnecessary gates, as they are part of the optimizable circuit. 

One additional feature of the data-weaved kernel is the fact that the data-weaved models seem to have steeper curves with alignment during training, which is an indication of the trainability of the model. This could be interesting future work on the topic, to see whether these models are also capable of learning the optimal feature map quicker than the data-last counterparts.

While QEKs are not the most common form of quantum kernels, it is also vital to see how these results affect models outside of QEKs. As the use of Loschmidt-echo test or swap test is widespread in the quantum kernel landscape, this begs the question whether similar problems with gate ordering could be influencing results across the field of research. These issues tend to be difficult to notice, as even something as dire as the gate erasure bug does not ruin one's result, but only increases the costs of computational resources. 

\section{Conclusion}
In this study, we explored different models of layer architectures for quantum embedded kernels, focusing on the intricacies of gate sequencing and their impact on performance and trainability. Our research indicates that the data-first or data-last approaches lead to inefficiencies, while the models where the parameterized parts are put in between feature encoding seem to fare better in most metrics. These inefficiencies culminated in a situation where the nullification of non-feature-dependent gates by the gate erasure bug causes data-first kernels to be computationally equivalent to the data-weaved kernels.

Through comprehensive testing across four datasets, we demonstrated that the data-weaved architecture, which positions data embeddings at the ends of each layer, not only improved the accuracy in a majority of cases but also optimized the training process. The new architecture showed indications of a marked improvement in kernel alignment, suggesting a more efficient utilization of quantum resources. 

While adding additional layers can compensate for inherent architectural inefficiencies, we show how architectural choices can be used to gain advantages across the board. We highlight the importance of strategic gate placement and the active contribution of all parameters in the circuit. This study underscores the need for careful consideration in quantum circuit design, particularly in the NISQ-era, where every computational element counts.

\section*{Acknowledgment}

Additional thanks to Valter Uotila and Urvi Sharma for valuable comments about the readability of the article.

\bibliographystyle{IEEEtran}
\bibliography{conference_101719}

\appendix
\subsection{Ansatz structure}
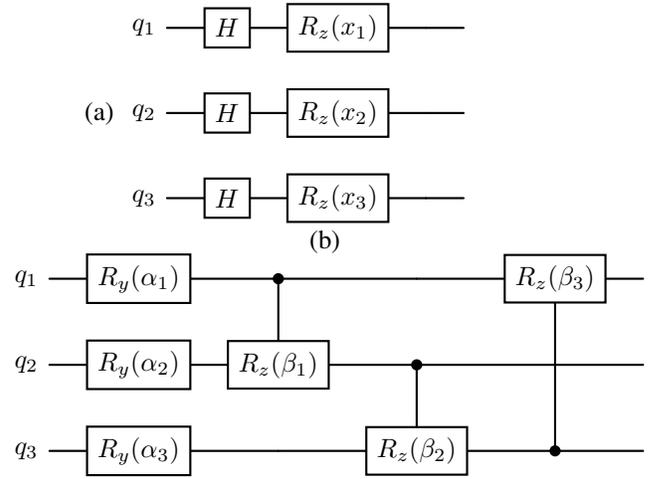
\begin{figure}[h!]
    \centering
    (a)
    \begin{quantikz}
        \lstick{$q_1$} & \gate{H} & \gate{R_z(x_1)} & & \\
        \lstick{$q_2$} & \gate{H} & \gate{R_z(x_2)} & & \\
        \lstick{$q_3$} & \gate{H} & \gate{R_z(x_3)} & & 
    \end{quantikz}
    \newline
    (b)
    \begin{quantikz}
        \lstick{$q_1$} & \gate{R_y(\alpha_1)} & \ctrl{1} & & \gate{R_z(\beta_3)} & \\
        \lstick{$q_2$} & \gate{R_y(\alpha_2)} & \gate{R_z(\beta_1)} & \ctrl{1} & & \\
        \lstick{$q_3$} & \gate{R_y(\alpha_3)} & & \gate{R_z(\beta_2)} & \ctrl{-2} & \\
    \end{quantikz}
    \caption{(a) The feature map (b) The ansatz using ring entanglement}
    \label{fig:kernel-circuit}
\end{figure}

The quantum circuit design for each architecture involves starting every data embedding layer with Hadamard gates followed by $R_z(X_n)$ rotations where $X$ is a data point from a dataset, the parameterized layer comprises of parameterized rotations $R_y(\alpha_n)$ followed by parameterized CZs: $CR_z(\beta_{n})$, where $\alpha$ and $\beta$ are trainable parameters. The $CR_z(\beta_{n})$ gates are applied in a circular pattern as seen in \autoref{fig:kernel-circuit} (b).

\end{document}